\documentclass[trackchanges,twocolumn]{aastex7}
\usepackage{longtable}
\usepackage{placeins}
\usepackage{threeparttable}
\usepackage{scalerel}
\usepackage{amsmath}
\usepackage{float}
\usepackage{graphicx}
\usepackage{mathptmx}
\usepackage{txfonts}
\usepackage[T1]{fontenc}
\usepackage{ae,aecompl}
\usepackage{graphicx}
\usepackage{capt-of}
\usepackage{lipsum}

\newcommand\HI{H\protect\scaleto{$I$}{1.2ex}}

\newcommand{\kms}{\mbox{km\,s$^{-1}$}}

\begin{document}

\title{Evolution of low surface brightness ultra-thin galaxies: The role of dark matter halo and bar formation on disk thickness}
\author[orcid=0000-0002-1250-4359,sname='K. Aditya']{K. Aditya}
\affiliation{ Raman Research Institute, C. V. Raman Avenue, 5th Cross Road, Sadashivanagar, Bengaluru, 560080, INDIA}\email{kaditya.astro@gmail.com}
\affiliation{Indian Institute of Astrophysics, Koramangala, Bengaluru 560 034, INDIA}
\author[sname='Arunima Banerjee']{Arunima Banerjee}
\affiliation{Department of Physics, Indian Institute of Science Education and Research (IISER), Tirupati - 517507, India}
\email{arunima@iisertirupati.ac.in}

\begin{abstract}
We investigate how stellar disks sustain their ultrathin structure throughout their evolution. We follow the evolution of ultrathin stellar disks 
with varying dark matter (DM) halo concentration ($c$) using collisionless $N$-body simulations with \texttt{AREPO}. We test models embedded in 
steep ($c = 12$), shallow ($c = 2$), and intermediate ($c = 6$) DM concentrations. Our models match the observed structural properties of the stellar
disk in the low surface brightness (LSB) ultrathin galaxy FGC~2366, specifically its surface brightness, disk scalelength, and vertical thinness 
($h_{z}/R_{D} = 0.1$), while excluding gas, allowing us to isolate the effects of DM. The internal disk heating mechanism driven by bars is suppressed 
in the LSB ultrathin stellar disks regardless of the DM concentration. The ratio of disk thickness ($h_z$) to scalelength ($R_D$) remains constant at
$\leq 0.1$ throughout their evolution. To clearly establish that the LSB nature of stellar disks is the key to preventing disk thickening, we construct 
the initial conditions by increasing the stellar mass fraction from $f_{s} \sim 0.01$ to $0.02$ and $0.04$, respectively, while keeping the total mass equal to
$10^{11} M_\odot$ and $h_z/R_D \leq 0.1$ unchanged. We find that models with a higher stellar mass fraction embedded in a shallow DM potential ($c = 2$) 
form bars and undergo significant disk thickening ($h_{z}/R_{D} \gg 0.1$) concurrent with the bar growth. We conclude that if the LSB disks are thin to begin 
with, they remain so throughout their evolution in isolation, regardless of the concentration of the DM halo.
\end{abstract}

\keywords{\uat{Galaxies}{573} --- \uat{Galaxy dynamics}{591} --- \uat{Galaxy kinematics}{602}  --- \uat{Gravitational instability}{668} --- \uat{Low surface brightness galaxies}{940} --- \uat{Galaxy bars}{2364}}

\section{Introduction}
Ultrathin galaxies (UTGs) are late-type edge-on disk galaxies characterized by a large major-to-minor-axis ratio $(a/b>10)$ 
\citep{goad1981spectroscopic,matthews1999extraordinary,aditya2022h, aditya2023h}. The UTGs lack a discernible bulge component 
and exhibit low surface brightness, with central surface brightness in B-band, $\mu_{B}>22.5\, mag\,arcsec^{-2}$ \citep{mcgaugh1996number,bothun1997low}. 
Although ubiquitous in observations\citep{1999BSAO...47....5K,kautsch2006catalog,bizyaev2017very, 2021ApJ...914..104B}, UTGs are notably rare in
$\Lambda CDM$ simulations. \cite{haslbauer2022high}, measured the sky-projected aspect ratio distribution in the $\Lambda CDM$ simulations:  
IllustrisTNG \citep{vogelsberger2014introducing, pillepich2018simulating}, and EAGLE \citep{2015MNRAS.446..521S} and found that these 
simulations are deficient in galaxies with intrinsically thin disks. Additionally, \cite{2017MNRAS.467.2879B} show that 
these simulations are deficient in bulge-dominated low-mass galaxies. However, more recent studies by \cite{hu2024formation} and 
\cite{xu2024illustristng} show that there is no shortage of thin disc galaxies in TNG-50. It is now well established that the bars, 
spiral arms \citep{saha2014disc, aumer2016age, grand2016vertical} can significantly heat the stellar disks radially, and 
Giant molecular clouds \citep{jenkins1990spiral} can heat the stellar disks radially and isotropically. However, the minimal 
vertical thickness indicates a negligible effect of disk heating agents in UTGs. Thus, the formation, evolution, and sustenance 
of these extremely thin stellar disks in UTGs remain a mystery and challenge our current understanding of galaxy formation and 
evolution models.

UTGs are distinguished by their extreme structural and photometric properties and characterized by equally extreme 
kinematic and dynamical features. \cite{komanduri2020dynamical,10.1093/mnras/stab155} have shown UTGs have central 
vertical velocity dispersion $(\sigma_{0s}=10\kms - 18\kms)$ comparable to the thin stellar disk in the Milky Way 
\citep{sharma2014kinematic}.  Additionally, UTGs are dynamically stable against growth of axisymmetric instabilities 
\citep{10.1093/mnras/stab155,aditya2022h,aditya2023h}, which may explain the low star formation rates observed in 
these galaxies \citep{narayanan2021star}. Furthermore, UTGs have a higher specific angular momentum compared 
to the ordinary spiral galaxies \citep{jadhav2019specific, aditya2022h, aditya2023h}, indicating that these 
galaxies rotate faster than ordinary spiral galaxies for a given stellar mass. It has been demonstrated that 
the extreme flattening of UTGs can be attributed to a compact and dense DM halo \citep{banerjee2013some}. 
This finding is further supported by DM models based on \HI{} 21 cm synthesis observations of two of the 
thinnest known galaxies, FGC 1440 (a/b=20) and FGC 2366 (a/b=22) \citep{aditya2022h, aditya2023h}. These results 
suggest that the DM halo plays an important role in regulating the vertical structure of galaxies hosting extremely thin stellar disks. 

Previous studies have shown that bar formation in low surface brightness (LSB) galaxies is 
closely tied to the mass distribution of the stellar and DM components. Numerical simulations of low surface brightness galaxies indicate that bar formation requires a disk mass at least twice as high as typically inferred for LSBs \citep{mayer2004formation}. Disks with low surface density are generally stable against bar formation, especially within halos with relatively higher DM concentration. Similarly, simulations by \cite{2001ApJ...546..176S} show that a stellar disk embedded in a DM halo with a large core radius is stable to $m=2$ modes \citep{toomre1981amplifies}. This is 
further corroborated in the numerical work by \cite{ghosh2014suppression}, who show that a dominant DM 
halo can suppress the formation of both axisymmetric and non-axisymmetric instabilities like bars and spiral formation 
in the prototypical low surface brightness UTG UGC 7321. Besides, cosmological simulations show that LSBs 
often inhabit halos with relatively low central density \citep{bailin2005concentrated} or high spin 
\citep{perez2022formation, chim2025study}, yielding extended stellar disks with low central densities. \cite{chim2025study} shows that the lower bar fraction in LSBs in the TNG-100 simulations is associated with 
their higher spin and gas content, factors known to inhibit bar formation and growth.

While previous studies have addressed the influence of DM on the vertical structure of ultra-thin galaxies (UTGs) 
using semi-analytic models, and have probed the connection between bar formation and DM concentration in low surface 
brightness (LSB) galaxies, these investigations have largely treated these effects in isolation. The nexus between the 
shape of the DM halo and the surface brightness of the stellar disk, and how they jointly govern the bar 
formation and ultimately regulate the vertical structure of low surface brightness UTGs hitherto remain unexplored.

In the present study, we will perform a series of N-body simulations using AREPO \citep{weinberger2020arepo} to investigate: \textit{how stellar disks sustain their ultrathin structure throughout their evolution ?}. We will 
describe the initial conditions in \S 2 and present the results from the analysis of the N-body simulations in \S 3. 
Finally, we will conclude in \S4.

\section{Initial conditions}
We use FGC 2366 as a template for constructing the initial conditions corresponding to the various models of UTGs presented in the study. FGC 2366 is the thinnest known galaxy 
with an extraordinarily large major-to-minor axis ratio $(a/b\approx 22)$. The stellar photometry, total rotation curve, and DM models for FGC 2366 are available in \cite{aditya2023h}. We 
construct the initial condition corresponding to FGC 2366 using the Disk Initial Conditions Environment 
(DICE) \citep{perret2016dice}. The structural parameters of the stellar disk are presented in Table 1. 
The stellar surface density is given by
\begin{equation}
 \Sigma_s (R,z) = \Sigma_{0} e^{-\,\left(\frac{R}{R_D}\right)} e^{-\,\left(\frac{z}{h_{z}}\right)},    
\end{equation}
where $\Sigma_{0}$ is the central stellar surface density, $R_{D}$ and $h_{z}$ are 
the exponential stellar disk scalelength and disk scaleheight, respectively.
The DM is modeled using Navarro-Frenk-White (NFW) \citep{navarro1997universal} profile given by
\begin{equation}
\rho_{DM}(R)=\frac{\rho_{0}}{\left(1+\frac{R}{R_{s}}\right)^{2}\left(\frac{R}{R_{s}}\right)},
\end{equation}
where $\rho_0$ is the central density and $R_{s}$ the scalelength of DM halo. The rotation curve due to the NFW DM density is given by
\begin{equation}
 V(R)=V_{200} \sqrt{\frac{\rm{ln}(1+cx) -cx/(1+cx)}{x[\rm{ln}(1+c) -c/(1+c)]}}
\end{equation}
where $x=R/R_{200}$, $R_{200}$ is the radius at which the mean density of the DM halo is 200 
times the critical density. $V_{200}$ is the rotation velocity at $R_{200}$ and is equal to $0.73R_{200}$. 
The concentration parameter is defined as $c=R_{200}/R_{s}$. The value of $R_{200}$ and $c$ completely 
specify the NFW DM distribution. We present the values of the concentration parameter and $R_{200}$ 
used for constructing initial conditions in Table 2. We use the stellar density parameters from Table 1 and 
the DM halo parameters from Table 2 to construct the initial conditions that reproduce the observed 
properties of FGC 2366, referred to as \texttt{c6} in our study. The surface density of the observed model 
\texttt{c6} corresponds to a central surface brightness of $22.8\,\mathrm{mag\,arcsec^{-2}}$ in z-band. We 
derive the mass-to-light ratio in the z-band in order to convert the surface brightness in $L_{\odot}/pc^{2}$ 
to the surface density in $M_{\odot}/pc^{2}$ using the calibration given in \cite{bell2003optical}. See 
\cite{aditya2023h} for detailed optical photometry of FGC 2366. The empirical calibration between the color 
and the mass-to-light ratio in a given band is given as  $\log_{10}(M/L)=a_{\lambda} +b_{\lambda} (Color)$. We use the g-z magnitudes $(g=15.44,\, z=14.76)$ to derive the color $(g-z=0.7)$ and the values of 
$a_{z} =-0.17$ and $b_{z}=0.32$ tabulated in \cite{2003ApJS..149..289B} and find mass-to-light ratio 
equal to $1.12$ in the $z$ band \citep{aditya2023h}. The central surface densities of the other models scale 
proportionally with their respective stellar mass fractions, see Table 2. It has been shown by 
\citep{jadhav2019specific,aditya2022h, aditya2023h} that UTGs have a higher specific angular momentum 
than ordinary spiral galaxies. We present the values of specific angular momentum for different models in Table 2. Unlike ordinary spiral galaxies, FGC 2366 has a higher specific angular momentum for a given stellar 
mass \citep{aditya2023h}. Similarly, \cite{perez2022formation} found that LSBs in TNG-100 simulations 
have a higher specific angular momentum than ordinary spirals.

To investigate the role of DM in determining disk thickness, we construct two additional models: 
one with a lower DM concentration (\texttt{c2}), resulting in a shallower rotation curve compared to 
\texttt{c6}, and another with a higher central DM concentration (\texttt{c12}), which produces a steeper 
rotation curve. To clearly establish that low surface brightness is key for preventing disk thickening, we construct 
initial condition by increasing the stellar mass fraction from $f_{s}\sim 0.008$ corresponding to the LSB thin disk 
to  $0.02$ labeled as (\texttt{c2-M1}, \texttt{c6-M1}, and \texttt{c12-M1}) and $0.04$ labeled as(\texttt{c2-M2}, 
\texttt{c6-M2}, and \texttt{c12-M2}) respectively. In all the models, we keep the total mass fixed at $10^{11}M_{\odot}$. 
Also, we ensure that all our initial conditions comply with the thin disk criterion $h_{z}/R_{D} \leq 0.1$, despite the variation 
in the DM concentration and stellar mass fraction. All the models have been initialized with $10^{6}$ stellar and dark 
matter particles, typical for simulations studying bar formation and disk thickening \citep{sellwood2013relaxation, sellwood2020three}. 
We present the initial conditions for our model galaxies in Figure 1 and Table 2. Furthermore, all models in our study are purely 
collisionless. Previous studies have shown that gas generally inhibits bar formation by facilitating angular 
momentum exchange between the stellar and gaseous components \citep{athanassoula2003determines, masters2012galaxy, sodi2017low}. 
Thus, using collisionless N-body simulations represents a conservative approach for assessing the role of bar instabilities on the 
vertical thickness of stellar disks. \\

\begin{table}
\caption{Structural parameters derived from optical photometry of FGC 2366.}
\hspace*{-2cm}
\begin{tabular}{|l|c|c|c|c|}
\hline
Parameters& $\Sigma_{0}$ & $R_{D}$&$h_{z}$ & a/b   \\
Galaxy         &$M_{\odot}$ $pc^{-2}$&kpc&kpc&\\
\hline    
\hline
FGC 2366 &24.1  &2.6    &0.29   &  21.6                      \\
\hline\end{tabular}
\end{table}

\begin{table*}
\caption{Initial conditions for the simulations}
\begin{tabular}{|l|c|c|c|c|c|c|c|}
\hline
Model                 &  $ N_{s}= N_{DM}$       & $f_{ s}$ & $\log(M/M_{\odot})$ & c &$R_{200}$&$\log(j_{*}/kpc\,kms^{-1})$& $\log(\Sigma_{0}/M_{\odot}kpc^{-2})$    \\ 
\hline    
\hline 
\texttt{c2}           &   $10^{6}$      &  0.008   &  11.04 &2.1&70.8  &2.5&7.6    \\
\texttt{c6}           &   $10^{6}$      &  0.008   &  10.95 &6  &62.8  &2.6&7.3    \\
\texttt{c12}          &   $10^{6}$      &  0.008   &  11.02 &12 &67    &2.8&7.4    \\
\texttt{c2-M1}        &   $10^{6}$      &  0.02    &  11.18 &2  &83.2  &2.6&8.1    \\
\texttt{c6-M1}        &   $10^{6}$      &  0.02    &  10.95 &6  &62.5  &2.7&7.7    \\
\texttt{c12-M1}       &   $10^{6}$      &  0.02    &  10.99 &12 &65    &2.8&7.8    \\
\texttt{c2-M2}        &   $10^{6}$      &  0.04    &  11.16 &2  &80.4  &2.7&8.4    \\
\texttt{c6-M2}        &   $10^{6}$      &  0.04    &  10.95 &6  &62    &2.7&8.0    \\
\texttt{c12-M2}       &   $10^{6}$      &  0.04    &  10.99 &12 &64    &2.8&8.1    \\
\hline
 \end{tabular}
\end{table*}

\begin{figure*}
\begin{tabular}{ccc}
\resizebox{180mm}{40mm}{\includegraphics{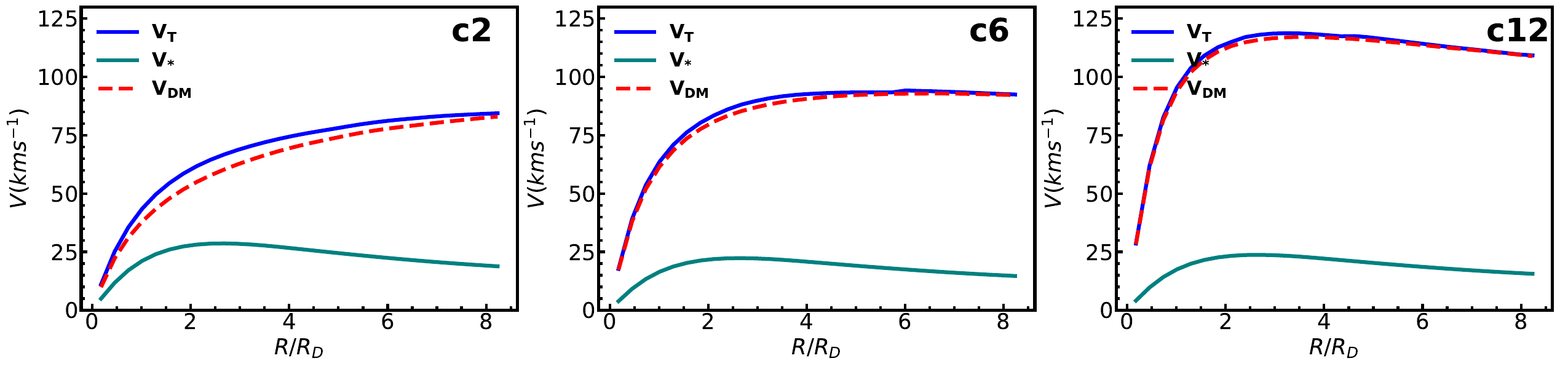}}\\
\resizebox{180mm}{40mm}{\includegraphics{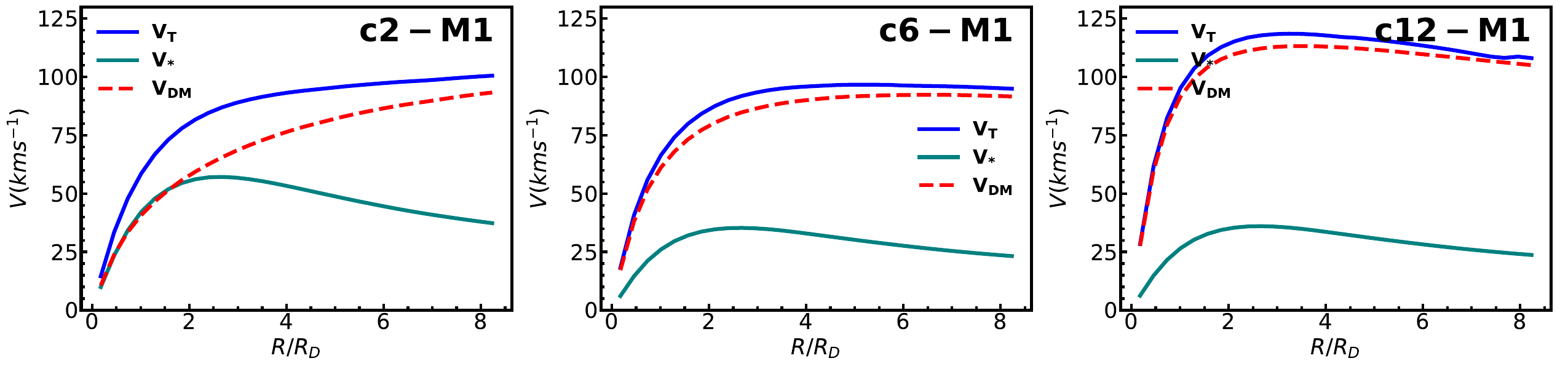}}\\
\resizebox{180mm}{40mm}{\includegraphics{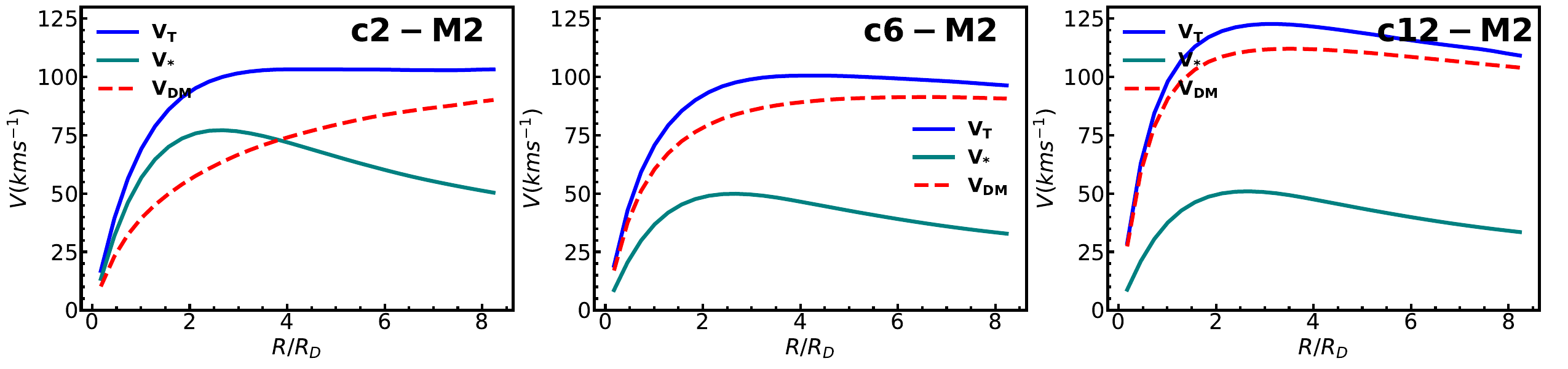}}\\
\end{tabular}
\caption{Initial conditions corresponding to various thin disk models 
in our study. All the models have a total mass equal to $10^{11}M_{\odot}$ and 
$h_{z}/R_{D}\leq 0.1$. The models in the first row with labels \texttt{c2}, \texttt{c6}, \texttt{c12}
have stellar mass fraction equal to $0.008$ and DM concentration equal to 2, 6, and 12, respectively. 
The model \texttt{c6} matches the observed properties of the thinnest known galaxy, FGC 2366. The models 
in the second row with labels \texttt{c2-M1}, \texttt{c6-M1}, \texttt{c12-M1} have a higher stellar mass 
fraction equal to $0.02$ and a DM halo concentration indicated by their labels. The models 
in the third row, labeled as \texttt{c2-M2}, \texttt{c6-M2}, and \texttt{c12-M2} have the 
highest stellar mass fraction among our models, equal to $0.04$.}  
\end{figure*}

\section{Results and discussion}
\begin{figure*}
\resizebox{170mm}{220mm}{\includegraphics{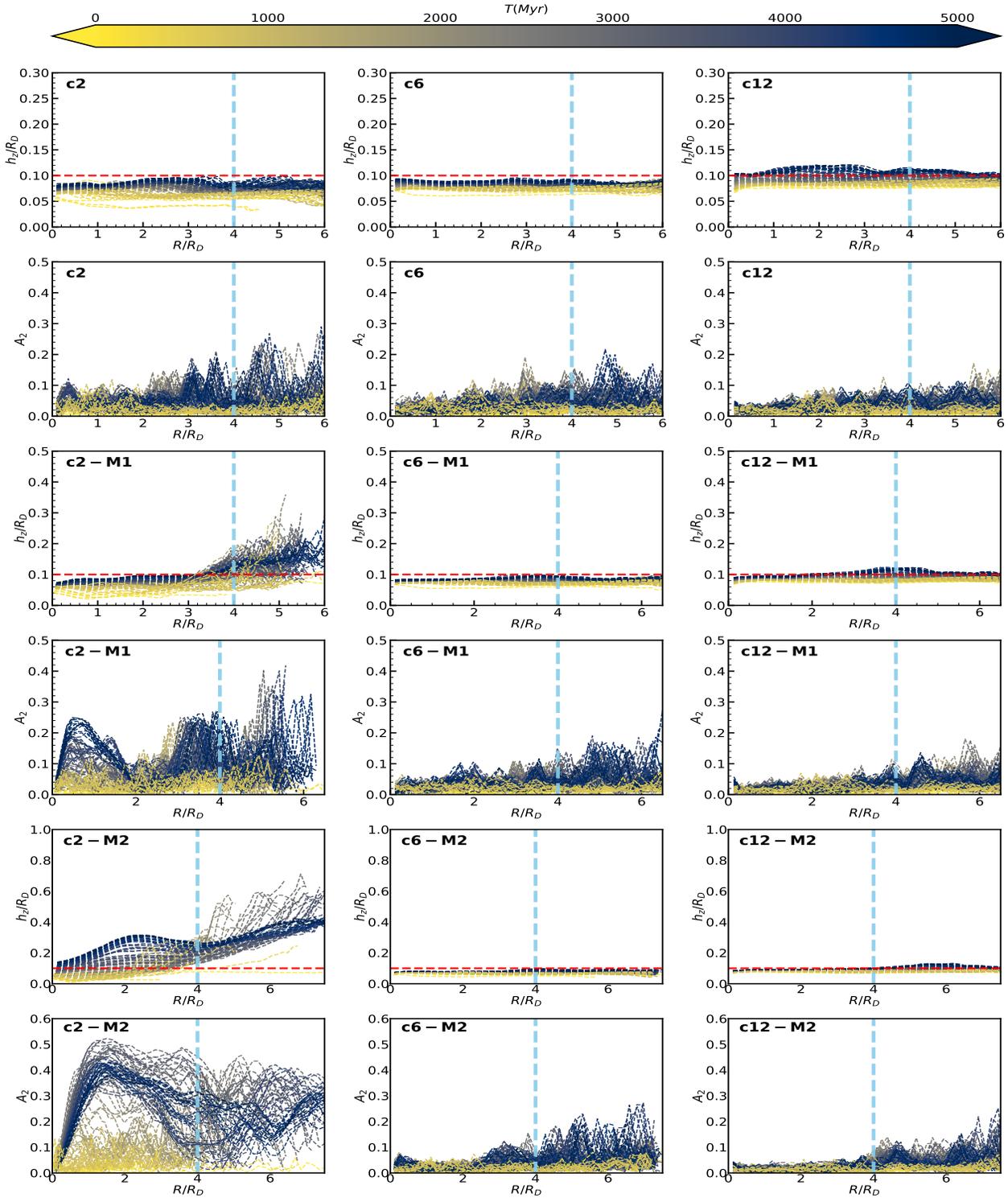}}\\
\caption{The time evolution of $h_{z}/R_{D}$ and the amplitude of the $m=2$ modes for different models presented 
in our work. In the top two panels, we show the models \texttt{c2}, \texttt{c6}, and \texttt{c12}, and in the following 
panels, we show their massive counterparts with a higher stellar mass fraction. We indicate the typical extent of the 
stellar disk with a vertical dashed line at $4R_{D}$. The horizontal line indicates the thin disk criterion; $h_{z}/R_{D}=0.1$.}
\end{figure*}

\begin{figure*}
\resizebox{190mm}{240mm}{\includegraphics{./img/Unified2.pdf}}\\
\vspace*{-2.0cm}
\caption{The face-on and the edge-on orientation of the snapshots at $T=4.5Gyr$. We show the 
face-on and the edge-on view corresponding to \texttt{c2}, \texttt{c6}, and \texttt{c12} in the top 
two panels, respectively, and their massive counterparts with a higher stellar mass fraction 
in the following panels.}
\end{figure*}

\begin{figure*}
\resizebox{190mm}{65mm}{\includegraphics{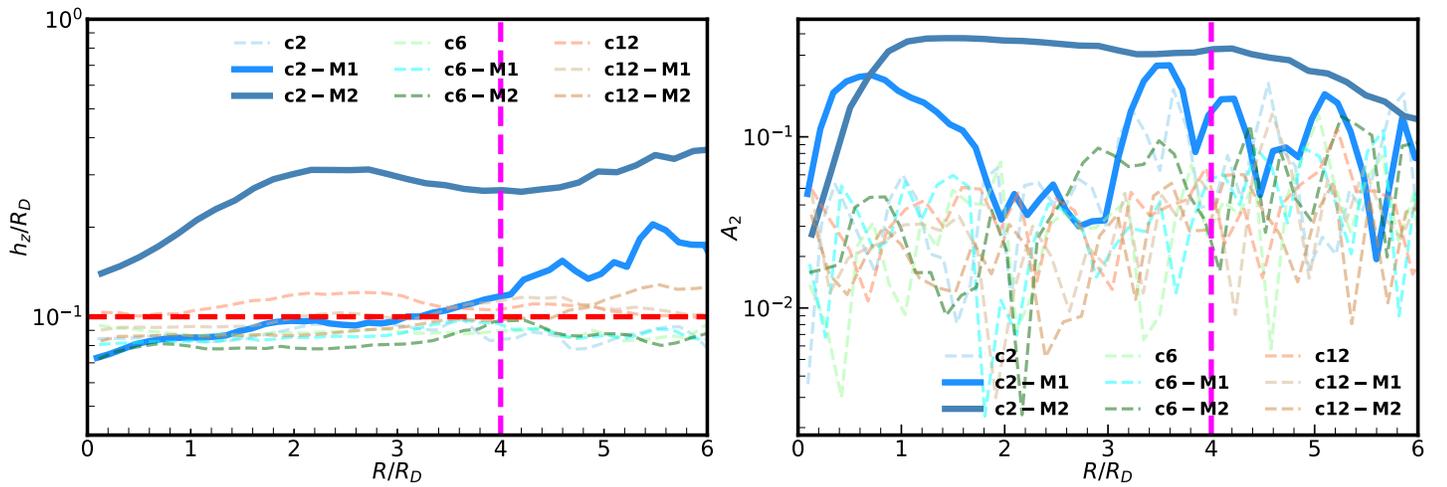}}
\caption{The values of $h_{z}/R_{D}$ and the Fourier amplitude of the $m=2$ mode at $T=5Gyr$ for 
different models. The solid lines represent the models \texttt{c2-M1} and \texttt{c2-M2}, which 
undergo significant disk heating concurrent with the growth of stellar bars. The other models in 
the study resist bar formation, thereby evading bar-driven disk thickening. The horizontal line 
indicates the thin disk criterion $(h_{z}/R_{D})=0.1$, and the vertical line represents 
$4R_{D}$, beyond which the surface density of the thin disk becomes negligible.}
\end{figure*}

We evolve the initial conditions for different models using publicly available code AREPO \citep{weinberger2020arepo}. We evolve our initial conditions for $5Gyrs$ and save 
the snapshot after every $50Myr$. We show the time evolution of $h_{z}/R_{D}$ corresponding to the 
different models of the ultrathin disks in Figure 2, and the face-on and edge-on projection of 
the surface density in Figure 3. We quantify the contribution of the internal disk 
heating due to bars by computing the Fourier amplitude of $m=2$ modes. The amplitude of the 
$n^{th}$ Fourier mode at radius $R$ is given by;
\begin{equation}
 a_{n}(R) = \sum_{i=1}^{N}m_{i}(R)\cos(n\theta_{i}),\\ 
 \quad   b_{n}(R) = \sum_{i=1}^{N}m_{i}(R)\sin(n\theta_{i}). 
\end{equation}
where, $n=0, \,1,\,2, ..$. In the above, $m_{r }(R)$ and $\theta_{i}$ are the mass and the azimuthal angle 
of the $i^{th}$ particle at radius $R$, and $N$ is the total number of particles at that radial position.
The strength of the $n^{th}$ mode is defined as;
\begin{equation}
 \frac{A_{n}}{A_{0}} = max\left(\frac{ \sqrt{a_{n}^{2} + b_{n}^{2} }}{\sum_{i=1}^{N}m_{i}}\right).   
\end{equation}

Upon inspecting Figure 2, we find that $h_{z}/R_{D}$ for \texttt{c2, c6, c12} at $5Gyrs$ has 
hardly changed from its initial value equal to 0.1. Further, from the face-on projections shown 
in Figure 3 (row 1), we can see that the models \texttt{c2, c6, c12} do not form bars and that 
the disks remain thin regardless of the concentration of the DM halo. This highlights the fact that the low surface brightness galaxies do not have sufficient self-gravity to support bar 
formation. As a result, internal disk heating due to bars is largely suppressed in isolated LSBs, 
allowing them to maintain their ultrathin disk structure throughout their evolution Figure 3 (row 2). 
Thus, in the LSB models \texttt{c2}, \texttt{c6}, and \texttt{c12}, the shape of the DM halo has little impact on the vertical structure or stellar surface density.

In order to establish that the low surface brightness of the stellar disk is key, we construct 
another set of initial conditions by increasing the stellar mass fraction from $0.008$ to 
$0.02$ (\texttt{c2-M1, c6-M1, c12-M1})  and $0.04$ (\texttt{c2-M2, c6-M2, c12-M2}). We find 
that the models \texttt{c2-M1} and \texttt{c2-M2} not only form bars but show a significant 
increase in disk thickness $(h_{z}/R_{D}>0.1)$, see Figures 2 and 3. The thickness of the stellar disk in both these models flares with radius. Flaring in the stellar disk is a common feature in massive 
galaxies; for example, see \cite{sotillo2023disc} for flaring stellar disks in Milky Way analogs in TNG-50. Also, see \cite{1997A&A...320L..21D, ossa2023flares}, for observational evidence of flaring stellar disks. We compare the value of $h_{z}/R_{D}$ and $A_{2}$ at the end of $5\,Gyr$ 
for different models in our study in Figure 4. We can see from  Figure 4 that the model \texttt{c6-M1}, \texttt{c12-M1}, \texttt{c6-M2}, and \texttt{c12-M2} neither form bars 
nor show signatures of significant disk thickening $(h_{z}/R_{D}\nless0.1)$, despite higher 
self-gravity due to a higher stellar mass fraction. This indicates that the steeper DM potential inhibits bar formation and prevents disk thickening. Previous studies by \cite{banerjee2016mass} 
showed that thin disk galaxies are embedded in DM halos with steeply rising rotation curves or 
steep central DM potential. Our results suggest that the low surface brightness of the stellar disk, in combination with
a steep inner DM density profile, is critical for preventing bar-driven disk thickening and maintaining the 
ultrathin vertical structure. We also note the presence of faint concentric ring-like features in the face-on stellar surface density maps across all models shown in Figure~3.

\subsection{Stability of ultrathin stellar disks}
In \cite{10.1093/mnras/stab155,aditya2022h,aditya2023h}, we have shown that ultrathin galaxies are highly stable with a median stability higher than the nearby spiral galaxies \citep{romeo2013simple}. \cite{aditya2023stability,aditya2024does} using a sample of LSBs from the SPARC catalog \citep{lelli2016sparc} and linear perturbation analysis show that a rigid DM halo plays an important role in stabilizing LSB galaxies against axis-symmetric instabilities. Using N-body \cite{sellwood2016bar} showed that bar instabilities are quelled when the disk is immersed in a massive static DM halo. However, \citep{sellwood2016bar} also points out that a live halo encourages bar formation. More recent studies by \cite{jang2023effects} using N-Body simulations of Milky Way type galaxies showed that the bar formation critically depends on the central mass concentration ($CMC$) and the minimum values of the Toomre stability criterion ($Q_{min}$) \citep{toomre1964gravitational}, given by;
\begin{equation}
(Q_{min}/1.2)^{2}  + (CMC/0.05)^{2} \leq 1.     
\end{equation}
\cite{jang2023effects} in their study model the DM halo using 
a Hernquist profile \citep{hernquist1990analytical}, compared to the NFW DM profile used in our study. Furthermore, the galaxy models considered by \citet{jang2023effects} 
have stellar and DM components comparable in size and mass to those of the Milky Way, 
and include a varying central mass concentration. In comparison, our models represent LSBs, with structural properties that match those of FGC 2366, the thinnest known disk galaxy. \cite{karachentsev1993flat, matthew2000properties} 
show that UTGs by definition are bulgeless systems. Further, optical photometry by \cite{aditya2023h} reveals
that FGC 2366 does not host a stellar bulge. Since our UTG models are based on the structural parameters of FGC 2366, which lacks a 
central bulge component, the bar formation criterion suggested by \cite{jang2023effects} is just $Q_{min}\leq1.4$. We show the time evolution of $Q$ for our models of ultrathin galaxies in Figure 5. All the ultrathin LSB models which do not form bars; \texttt{c2}, \texttt{c6} and \texttt{c12} have $Q>1.4$. We find that the thin disk models \texttt{c2-M1} and \texttt{c2-M2}, which form bars and undergo disk thickening, have $Q_{min}<1.4$. Further, we note that models with steep inner DM halo have $Q_{min}>1.4$ and do not form bars. Indirectly, $Q_{min}>1.4$ suggests that UTGs in our study are not susceptible to disk thickening driven by bars.

\subsection{Efstathiou-Lake-Negroponte criterion}
\cite{efstathiou1982stability} investigated the  global stability of disk galaxies using N-body experiments and showed that the models with
\begin{equation}
    X= \frac{V_{max}}{\left(GM_{D}/R_{D}\right)^{1/2}} \leq 1.1,
\end{equation}
favor bar formation. In the above equation $V_{max}$, $M_{D}$, 
$R_{D}$ are the maximum circular velocity, the mass of the stellar disk, and the disk scalelength. We show  $X$ as a function of time for all our galaxies in Figure 6. We can see that all our models except \texttt{c2-M1} and \texttt{c2-M2} have $X>1.1$. The models \texttt{c2-M1} and \texttt{c2-M2} not only have $Q_{min}<1.4$ but also satisfy $X<1.1$. The models \texttt{c2-M1} and \texttt{c2-M2} have a higher 
stellar mass fraction and hence higher self-gravity, which supports bar formation. 
However, models with higher DM concentration do not form bars despite a higher stellar mass fraction. This occurs because when the DM concentration is lowered while keeping the total mass constant, the reduced central DM mass is compensated by a higher stellar mass in the central regions. Thus, the shallow DM potential does not provide sufficient centrifugal support to counteract the increased self-gravity of the stellar disk in the central regions, leading to  $X<1.1$ 
and eventually resulting in disk thickening driven by bar formation. However, the steeper DM potential provides significant centrifugal support and stabilizes the ultrathin stellar disk against bar formation.

\begin{figure*}
\resizebox{170mm}{140mm}{\includegraphics{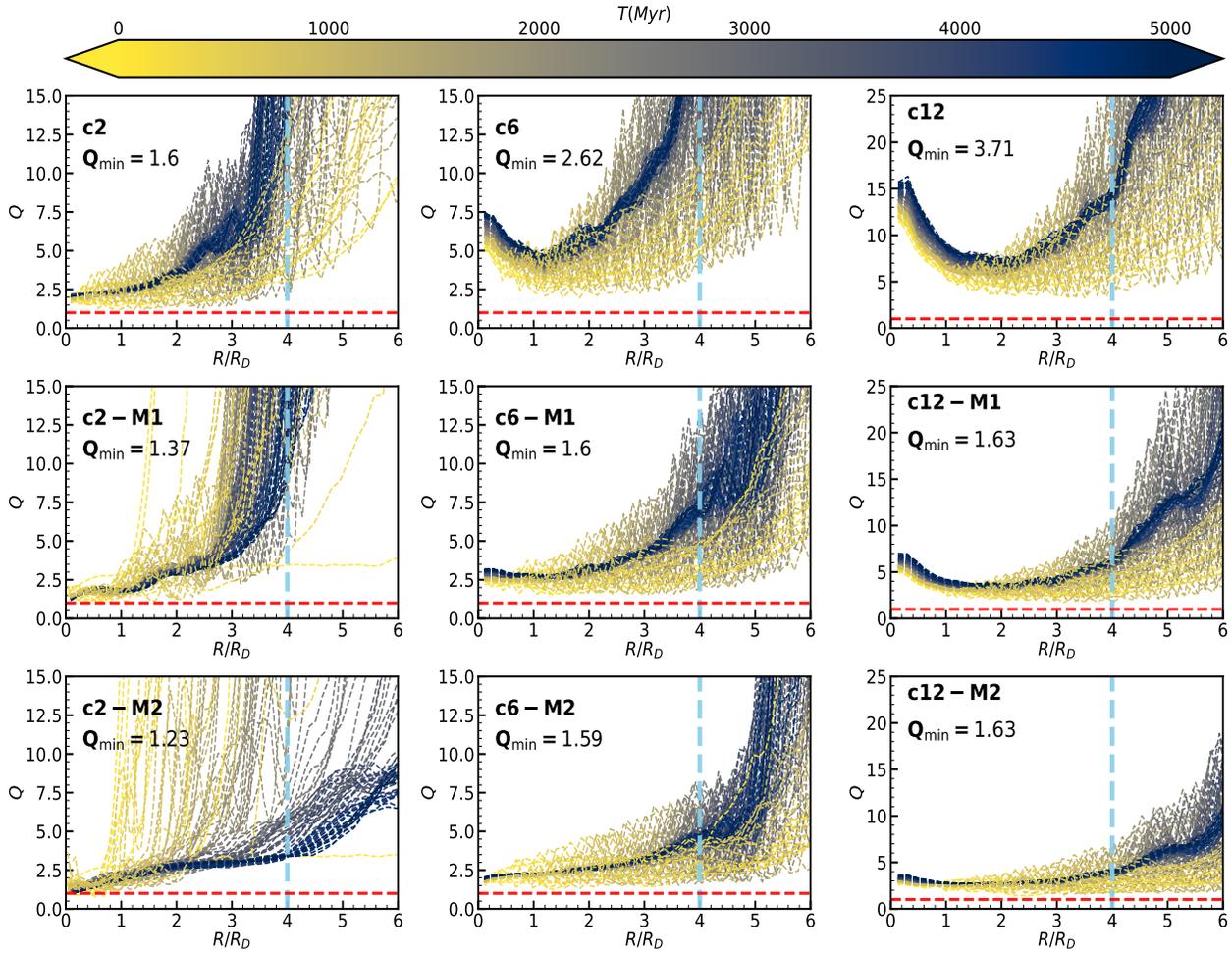}}\\
\caption{The time evolution of the stability criterion $Q$. In the top two panels, we show the models 
\texttt{c2}, \texttt{c6}, and \texttt{c12}, and in the following panels, we show their massive counterparts 
with a higher stellar mass fraction. We indicate the typical extent of the stellar disk with a vertical dashed 
line at $4R_{D}$. The horizontal line indicates the criterion for marginal stability; $Q=1$. }
\end{figure*}

\begin{figure*}
\resizebox{170mm}{120mm}{\includegraphics{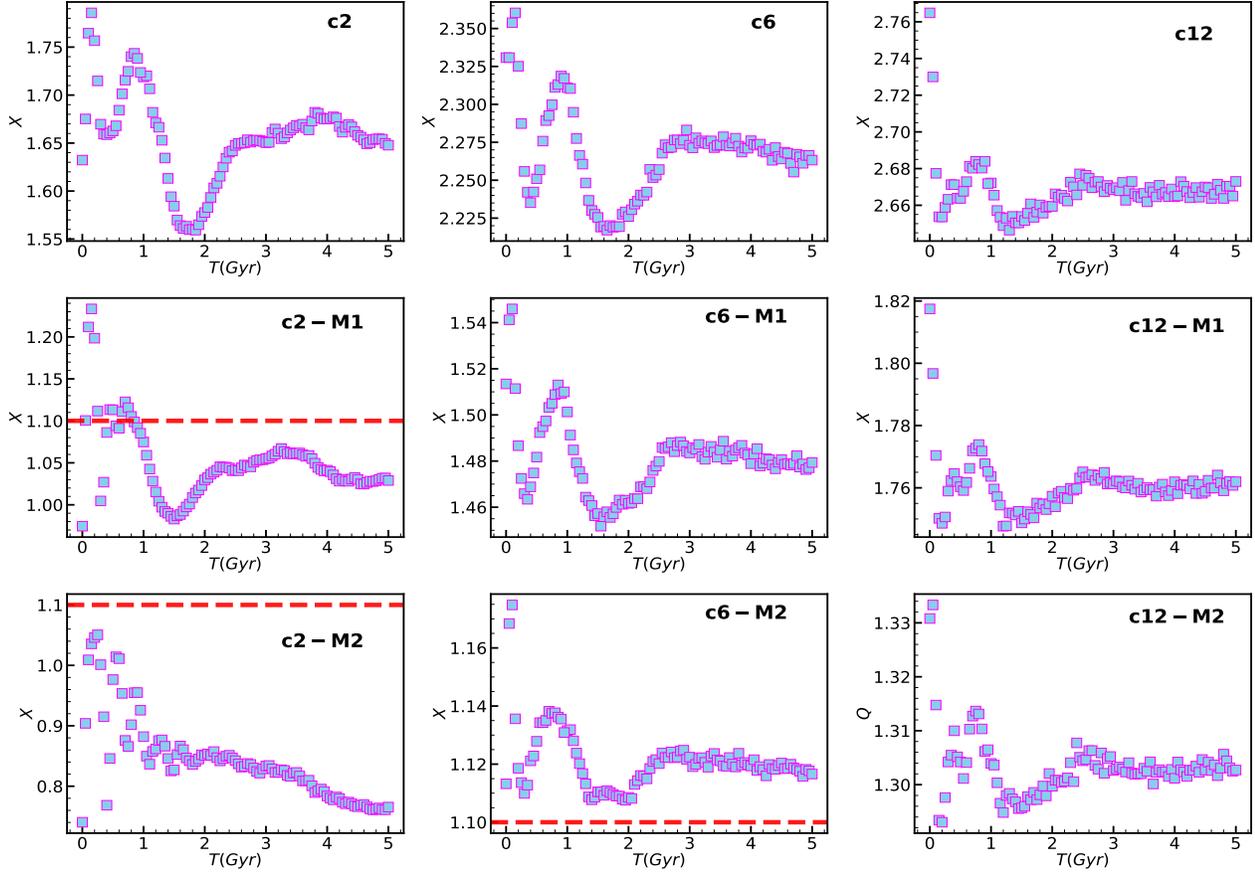}}\\
\caption{The time evolution of the bar instability criterion $X$. In the top two panels, we show the models 
\texttt{c2}, \texttt{c6}, and \texttt{c12}, and in the following panels, we show their massive counterparts 
with a higher stellar mass fraction. The horizontal line indicates the bar instability criterion $X=1.1$.}
\end{figure*}

\section{Conclusions}
The results presented in this work show that the collisionless models of low surface brightness ultrathin stellar disks do not form bars and that the disks remain thin regardless of the concentration of the DM halo. 
The initial disk scaleheight ratio, $h_{z}/R_{D}$, remains nearly constant, suggesting that the low surface brightness ultrathin stellar disks are highly resistant to internal heating mechanisms due to bars and spiral arms, which typically drive vertical heating and disk thickening \citep{saha2010effect, sellwood2020three, ghosh2024bars}.
Furthermore, upon increasing the mass fraction, we find that the ultrathin stellar disks embedded in a shallow DM halo undergo disk thickening concurrent with bar formation. In contrast, a steep DM halo stabilizes the ultrathin stellar disks against bar formation even with a higher stellar mass fraction. 

We note, however, that our models do not include gas and hence are not ideal analogues of FGC 2366. They are constructed to 
match the observed structural parameters of the galaxy, such as the stellar surface density, scalelength, and thickness of the 
stellar disk. Our models consider only the stellar disk embedded in a DM halo, since our aim here was to isolate the dynamical effect of DM halo concentration on bar formation and eventually on the 
vertical thickness of stellar disks.

Several studies have shown that the presence of gas can suppress bar formation. For instance, observational studies by 
\citet{masters2012galaxy}, \citet{sodi2017low}, and \citet{zhou2021correlation} suggest that angular momentum exchange 
between stars and gas inhibits bar growth \citep[see also][]{athanassoula2003determines}. Likewise, a recent study by 
\citet{ramirez2025study}, using TNG-100 simulations, shows that LSB galaxies with stellar mass below  $10^{11}M_{\odot}$ 
exhibit a consistently lower bar fraction than high surface brightness galaxies, which could be attributed to their higher gas fractions. So, although we have not included gas in our models, the available evidence suggests that doing so would have further inhibited bar formation, making our purely stellar models a conservative choice.

Our findings are consistent with previous simulations of low surface brightness galaxies. \cite{mayer2004formation} found that 
bar formation requires disk masses that are at least twice those inferred for LSBs. Similar to our results, they found that LSB disks 
are highly stable against bar formation for a wide array of halo parameters. Interestingly, \citet{ghosh2023bars, ghosh2024bars} have shown 
that massive, thick stellar disks can still form bars. These results highlight the role of the stellar-to-halo mass ratio and internal disk 
dynamics on bar formation in N-body simulations.

We also acknowledge that other factors, such as the shape of the DM halo (spherical or triaxial), and 
whether the halo is live or static, can all contribute to bar formation and influence the vertical thickness 
\citep{el2002dark, berentzen2006stellar}. For instance, \citet{narayanan2024does} propose that the quadrupole 
moment of the halo can drive long-lived spiral arms in LSB galaxies, while \citet{hu2016stellar} explore the role 
of triaxial DM halos on spiral formation in Milky Way like systems. .

We conclude that if the ultrathin LSBs are thin, they remain so throughout their evolution in isolation regardless 
of the concentration of the DM halo. The low surface brightness nature of these galaxies is paramount to their 
thin disk structure. Bar formation is suppressed in these galaxies, which inhibits the internal disk heating 
mechanism, ensuring that LSBs maintain their relatively thin stellar disks. In contrast, their massive 
thin disk counterparts with a higher stellar mass fraction residing in a shallow inner DM profile readily form 
bars and undergo disk thickening supported by the growth of stellar bars. Thus, our results emphasize that 
the low surface brightness nature of the thin stellar discs in combination with the steep inner DM halo plays a 
key role in suppressing the bar formation that can otherwise heat the stellar discs. 

\section{Acknowledgement}
We gratefully acknowledge the use of the high-performance computing facility ‘NOVA’ at the Indian Institute of Astrophysics, Bengaluru, 
India, where all simulations were carried out

\clearpage
\bibliographystyle{aasjournalv7}
\bibliography{2366_d2.bib}{}

\begin{thebibliography}{}
\expandafter\ifx\csname natexlab\endcsname\relax\def\natexlab#1{#1}\fi
\providecommand{\url}[1]{\href{#1}{#1}}
\providecommand{\dodoi}[1]{doi:~\href{http://doi.org/#1}{\nolinkurl{#1}}}
\providecommand{\doeprint}[1]{\href{http://ascl.net/#1}{\nolinkurl{http://ascl.net/#1}}}
\providecommand{\doarXiv}[1]{\href{https://arxiv.org/abs/#1}{\nolinkurl{https://arxiv.org/abs/#1}}}

\bibitem[{K. Aditya(2023)Aditya}]{aditya2023stability}
Aditya, K. 2023, \bibinfo{title}{Stability of galaxies across morphological
  sequence,} Monthly Notices of the Royal Astronomical Society, 522, 2543

\bibitem[{K. Aditya(2024)Aditya}]{aditya2024does}
Aditya, K. 2024, \bibinfo{title}{How does dark matter stabilize disc galaxies?}
  Monthly Notices of the Royal Astronomical Society, 532, 3839

\bibitem[{K. Aditya \& A. Banerjee(2021)Aditya \&
  Banerjee}]{10.1093/mnras/stab155}
Aditya, K., \& Banerjee, A. 2021, \bibinfo{title}{{How “cold” are the
  stellar discs of superthin galaxies?},} Monthly Notices of the Royal
  Astronomical Society, \dodoi{10.1093/mnras/stab155}

\bibitem[{K. Aditya {et~al.}(2023)Aditya, Banerjee, Kamphuis, Mosenkov,
  Makarov, \& Borisov}]{aditya2023h}
Aditya, K., Banerjee, A., Kamphuis, P., {et~al.} 2023, \bibinfo{title}{H i 21cm
  observations and dynamical modelling of the thinnest galaxy: FGC 2366,}
  Monthly Notices of the Royal Astronomical Society, 526, 29

\bibitem[{K. Aditya {et~al.}(2022)Aditya, Kamphuis, Banerjee, Borisov,
  Mosenkov, Antipova, \& Makarov}]{aditya2022h}
Aditya, K., Kamphuis, P., Banerjee, A., {et~al.} 2022, \bibinfo{title}{H i 21
  cm observation and mass models of the extremely thin galaxy FGC 1440,}
  Monthly Notices of the Royal Astronomical Society, 509, 4071

\bibitem[{E. Athanassoula(2003)Athanassoula}]{athanassoula2003determines}
Athanassoula, E. 2003, \bibinfo{title}{What determines the strength and the
  slowdown rate of bars?} Monthly Notices of the Royal Astronomical Society,
  341, 1179

\bibitem[{M. Aumer {et~al.}(2016)Aumer, Binney, \&
  Sch{\"o}nrich}]{aumer2016age}
Aumer, M., Binney, J., \& Sch{\"o}nrich, R. 2016, \bibinfo{title}{Age--velocity
  dispersion relations and heating histories in disc galaxies,} Monthly Notices
  of the Royal Astronomical Society, 462, 1697

\bibitem[{J. Bailin {et~al.}(2005)Bailin, Power, Gibson, \&
  Steinmetz}]{bailin2005concentrated}
Bailin, J., Power, C., Gibson, B.~K., \& Steinmetz, M. 2005,
  \bibinfo{title}{How Concentrated Are The Haloes Of Low Surface Brightness
  Galaxies In The Cold Dark Matter Model?} arXiv preprint astro-ph/0502231

\bibitem[{A. Banerjee \& D. Bapat(2017)Banerjee \& Bapat}]{banerjee2016mass}
Banerjee, A., \& Bapat, D. 2017, \bibinfo{title}{Mass modelling of superthin
  galaxies: IC5249, UGC7321 and IC2233,} Monthly Notices of the Royal
  Astronomical Society, 466, 3753

\bibitem[{A. Banerjee \& C.~J. Jog(2013)Banerjee \& Jog}]{banerjee2013some}
Banerjee, A., \& Jog, C.~J. 2013, \bibinfo{title}{Why are some galaxy discs
  extremely thin?} Monthly Notices of the Royal Astronomical Society, 431, 582

\bibitem[{E.~F. Bell {et~al.}(2003)Bell, McIntosh, Katz, \&
  Weinberg}]{bell2003optical}
Bell, E.~F., McIntosh, D.~H., Katz, N., \& Weinberg, M.~D. 2003,
  \bibinfo{title}{The optical and near-infrared properties of galaxies. I.
  Luminosity and stellar mass functions,} The Astrophysical Journal Supplement
  Series, 149, 289

\bibitem[{E.~F. {Bell} {et~al.}(2003){Bell}, {McIntosh}, {Katz}, \&
  {Weinberg}}]{2003ApJS..149..289B}
{Bell}, E.~F., {McIntosh}, D.~H., {Katz}, N., \& {Weinberg}, M.~D. 2003,
  \bibinfo{title}{{The Optical and Near-Infrared Properties of Galaxies. I.
  Luminosity and Stellar Mass Functions},} The Astrophysical Journal Supplement
  Series, 149, 289, \dodoi{10.1086/378847}

\bibitem[{I. Berentzen {et~al.}(2006)Berentzen, Shlosman, \&
  Jogee}]{berentzen2006stellar}
Berentzen, I., Shlosman, I., \& Jogee, S. 2006, \bibinfo{title}{Stellar bar
  evolution in cuspy and flat-cored triaxial CDM halos,} The Astrophysical
  Journal, 637, 582

\bibitem[{D. Bizyaev {et~al.}(2017)Bizyaev, Kautsch, Sotnikova, Reshetnikov, \&
  Mosenkov}]{bizyaev2017very}
Bizyaev, D., Kautsch, S., Sotnikova, N.~Y., Reshetnikov, V.~P., \& Mosenkov,
  A.~V. 2017, \bibinfo{title}{Very thin disc galaxies in the SDSS catalogue of
  edge-on galaxies,} Monthly Notices of the Royal Astronomical Society, 465,
  3784

\bibitem[{D. {Bizyaev} {et~al.}(2021){Bizyaev}, {Makarov}, {Reshetnikov},
  {Mosenkov}, {Kautsch}, \& {Antipova}}]{2021ApJ...914..104B}
{Bizyaev}, D., {Makarov}, D.~I., {Reshetnikov}, V.~P., {et~al.} 2021,
  \bibinfo{title}{{Spectral Observations of Superthin Galaxies},} The
  Astrophysical Journal, 914, 104, \dodoi{10.3847/1538-4357/abfb03}

\bibitem[{G. Bothun {et~al.}(1997)Bothun, Impey, \& McGaugh}]{bothun1997low}
Bothun, G., Impey, C., \& McGaugh, S. 1997,
  \bibinfo{title}{Low-surface-brightness galaxies: hidden galaxies revealed,}
  Publications of the Astronomical Society of the Pacific, 109, 745

\bibitem[{C. {Bottrell} {et~al.}(2017){Bottrell}, {Torrey}, {Simard}, \&
  {Ellison}}]{2017MNRAS.467.2879B}
{Bottrell}, C., {Torrey}, P., {Simard}, L., \& {Ellison}, S.~L. 2017,
  \bibinfo{title}{{Galaxies in the Illustris simulation as seen by the Sloan
  Digital Sky Survey - II. Size-luminosity relations and the deficit of
  bulge-dominated galaxies in Illustris at low mass},} Monthly Notices of the
  Royal Astronomical Society, 467, 2879, \dodoi{10.1093/mnras/stx276}

\bibitem[{K. Chim-Ramirez {et~al.}(2025)Chim-Ramirez, Cervantes-Sodi,
  Rosas-Guevara, P{\'e}rez-Monta{\~n}o, \& Bonoli}]{chim2025study}
Chim-Ramirez, K., Cervantes-Sodi, B., Rosas-Guevara, Y., P{\'e}rez-Monta{\~n}o,
  L.~E., \& Bonoli, S. 2025, \bibinfo{title}{Study of barred galaxies in
  IllustrisTNG100: the case of low surface brightness galaxies,} Monthly
  Notices of the Royal Astronomical Society, 539, 2262

\bibitem[{R. {de Grijs} \& R.~F. {Peletier}(1997){de Grijs} \&
  {Peletier}}]{1997A&A...320L..21D}
{de Grijs}, R., \& {Peletier}, R.~F. 1997, \bibinfo{title}{{The shape of galaxy
  disks: how the scale height increases with galactocentric distance.},} \aap,
  320, L21, \dodoi{10.48550/arXiv.astro-ph/9702215}

\bibitem[{G. Efstathiou {et~al.}(1982)Efstathiou, Lake, \&
  Negroponte}]{efstathiou1982stability}
Efstathiou, G., Lake, G., \& Negroponte, J. 1982, \bibinfo{title}{The stability
  and masses of disc galaxies,} Monthly Notices of the Royal Astronomical
  Society, 199, 1069

\bibitem[{A. El-Zant \& I. Shlosman(2002)El-Zant \& Shlosman}]{el2002dark}
El-Zant, A., \& Shlosman, I. 2002, \bibinfo{title}{Dark halo shapes and the
  fate of stellar bars,} The Astrophysical Journal, 577, 626

\bibitem[{S. Ghosh {et~al.}(2023)Ghosh, Fragkoudi, Di~Matteo, \&
  Saha}]{ghosh2023bars}
Ghosh, S., Fragkoudi, F., Di~Matteo, P., \& Saha, K. 2023, \bibinfo{title}{Bars
  and boxy/peanut bulges in thin and thick discs-II. Can bars form in hot thick
  discs?} Astronomy \& Astrophysics, 674, A128

\bibitem[{S. Ghosh {et~al.}(2024)Ghosh, Fragkoudi, Di~Matteo, \&
  Saha}]{ghosh2024bars}
Ghosh, S., Fragkoudi, F., Di~Matteo, P., \& Saha, K. 2024, \bibinfo{title}{Bars
  and boxy/peanut bulges in thin and thick discs-III. Boxy/peanut bulge
  formation and evolution in the presence of thick discs,} Astronomy \&
  Astrophysics, 683, A196

\bibitem[{S. Ghosh \& C.~J. Jog(2014)Ghosh \& Jog}]{ghosh2014suppression}
Ghosh, S., \& Jog, C.~J. 2014, \bibinfo{title}{Suppression of gravitational
  instabilities by dominant dark matter halo in low-surface-brightness
  galaxies,} Monthly Notices of the Royal Astronomical Society, 439, 929

\bibitem[{J.~W. Goad \& M.~S. Roberts(1981)Goad \&
  Roberts}]{goad1981spectroscopic}
Goad, J.~W., \& Roberts, M.~S. 1981, \bibinfo{title}{Spectroscopic observations
  of superthin galaxies,} Astrophysical Journal, Part 1, vol. 250, Nov. 1,
  1981, p. 79-86., 250, 79

\bibitem[{R.~J. Grand {et~al.}(2016)Grand, Springel, G{\'o}mez, Marinacci,
  Pakmor, Campbell, \& Jenkins}]{grand2016vertical}
Grand, R.~J., Springel, V., G{\'o}mez, F.~A., {et~al.} 2016,
  \bibinfo{title}{Vertical disc heating in Milky Way-sized galaxies in a
  cosmological context,} Monthly Notices of the Royal Astronomical Society,
  459, 199

\bibitem[{M. Haslbauer {et~al.}(2022)Haslbauer, Banik, Kroupa, Wittenburg, \&
  Javanmardi}]{haslbauer2022high}
Haslbauer, M., Banik, I., Kroupa, P., Wittenburg, N., \& Javanmardi, B. 2022,
  \bibinfo{title}{The High Fraction of Thin Disk Galaxies Continues to
  Challenge $\Lambda$CDM Cosmology,} The Astrophysical Journal, 925, 183

\bibitem[{L. Hernquist(1990)Hernquist}]{hernquist1990analytical}
Hernquist, L. 1990, \bibinfo{title}{An analytical model for spherical galaxies
  and bulges,} Astrophysical Journal, Part 1 (ISSN 0004-637X), vol. 356, June
  20, 1990, p. 359-364., 356, 359

\bibitem[{J. Hu {et~al.}(2024)Hu, Xu, \& Li}]{hu2024formation}
Hu, J., Xu, D., \& Li, C. 2024, \bibinfo{title}{Formation of Superthin Galaxies
  in IllustrisTNG,} Research in Astronomy and Astrophysics, 24, 075019

\bibitem[{S. Hu \& D. Sijacki(2016)Hu \& Sijacki}]{hu2016stellar}
Hu, S., \& Sijacki, D. 2016, \bibinfo{title}{Stellar spiral structures in
  triaxial dark matter haloes,} Monthly Notices of the Royal Astronomical
  Society, 461, 2789

\bibitem[{V. Jadhav~Y \& A. Banerjee(2019)Jadhav~Y \&
  Banerjee}]{jadhav2019specific}
Jadhav~Y, V., \& Banerjee, A. 2019, \bibinfo{title}{The specific angular
  momenta of superthin galaxies: Cue to their origin?} Monthly Notices of the
  Royal Astronomical Society, 488, 547

\bibitem[{D. Jang \& W.-T. Kim(2023)Jang \& Kim}]{jang2023effects}
Jang, D., \& Kim, W.-T. 2023, \bibinfo{title}{Effects of the Central Mass
  Concentration on Bar Formation in Disk Galaxies,} The Astrophysical Journal,
  942, 106

\bibitem[{A. Jenkins \& J. Binney(1990)Jenkins \& Binney}]{jenkins1990spiral}
Jenkins, A., \& Binney, J. 1990, \bibinfo{title}{Spiral heating of galactic
  discs,} Monthly Notices of the Royal Astronomical Society, 245, 305

\bibitem[{I. Karachentsev {et~al.}(1993)Karachentsev, Karachentseva, \&
  Parnovsky}]{karachentsev1993flat}
Karachentsev, I., Karachentseva, V., \& Parnovsky, S. 1993,
  \bibinfo{title}{Flat galaxy catalogue,} Astronomische Nachrichten, 314, 97

\bibitem[{I.~D. {Karachentsev} {et~al.}(1999){Karachentsev}, {Karachentseva},
  {Kudrya}, {Sharina}, \& {Parnovskij}}]{1999BSAO...47....5K}
{Karachentsev}, I.~D., {Karachentseva}, V.~E., {Kudrya}, Y.~N., {Sharina},
  M.~E., \& {Parnovskij}, S.~L. 1999, \bibinfo{title}{{The revised Flat Galaxy
  Catalogue.},} Bulletin of the Special Astrophysics Observatory, 47, 5.
\newblock \doarXiv{astro-ph/0305566}

\bibitem[{S. Kautsch {et~al.}(2006)Kautsch, Grebel, Barazza, \&
  Gallagher}]{kautsch2006catalog}
Kautsch, S., Grebel, E., Barazza, F., \& Gallagher, J. 2006, \bibinfo{title}{A
  catalog of edge-on disk galaxies-From galaxies with a bulge to superthin
  galaxies,} Astronomy \& Astrophysics, 445, 765

\bibitem[{A. Komanduri {et~al.}(2020)Komanduri, Banerjee, Banerjee, \&
  Sengupta}]{komanduri2020dynamical}
Komanduri, A., Banerjee, I., Banerjee, A., \& Sengupta, S. 2020,
  \bibinfo{title}{Dynamical modelling of disc vertical structure in superthin
  galaxy ‘UGC 7321’in braneworld gravity: an MCMC study,} Monthly Notices
  of the Royal Astronomical Society, 499, 5690

\bibitem[{F. Lelli {et~al.}(2016)Lelli, McGaugh, \& Schombert}]{lelli2016sparc}
Lelli, F., McGaugh, S.~S., \& Schombert, J.~M. 2016, \bibinfo{title}{SPARC:
  mass models for 175 disk galaxies with Spitzer photometry and accurate
  rotation curves,} The Astronomical Journal, 152, 157

\bibitem[{K.~L. Masters {et~al.}(2012)Masters, Nichol, Haynes, Keel, Lintott,
  Simmons, Skibba, Bamford, Giovanelli, \& Schawinski}]{masters2012galaxy}
Masters, K.~L., Nichol, R.~C., Haynes, M.~P., {et~al.} 2012,
  \bibinfo{title}{Galaxy Zoo and ALFALFA: atomic gas and the regulation of star
  formation in barred disc galaxies,} Monthly Notices of the Royal Astronomical
  Society, 424, 2180

\bibitem[{L. Matthew {et~al.}(2000)Matthew, van Driel, \&
  Gallagher}]{matthew2000properties}
Matthew, L., van Driel, W., \& Gallagher, J. 2000, \bibinfo{title}{Properties
  of" superthin" galaxies,} Building Galaxies; from the Primordial Universe to
  the Present, 107

\bibitem[{L. Matthews {et~al.}(1999)Matthews, Gallagher~III, \&
  Van~Driel}]{matthews1999extraordinary}
Matthews, L., Gallagher~III, J., \& Van~Driel, W. 1999, \bibinfo{title}{The
  Extraordinary “Superthin” Spiral Galaxy UGC 7321. I. Disk Color Gradients
  and Global Properties from Multiwavelength Observations,} The Astronomical
  Journal, 118, 2751

\bibitem[{L. Mayer \& J. Wadsley(2004)Mayer \& Wadsley}]{mayer2004formation}
Mayer, L., \& Wadsley, J. 2004, \bibinfo{title}{The formation and evolution of
  bars in low surface brightness galaxies with cold dark matter haloes,}
  Monthly Notices of the Royal Astronomical Society, 347, 277

\bibitem[{S.~S. McGaugh(1996)McGaugh}]{mcgaugh1996number}
McGaugh, S.~S. 1996, \bibinfo{title}{The number, luminosity and mass density of
  spiral galaxies as a function of surface brightness,} Monthly Notices of the
  Royal Astronomical Society, 280, 337

\bibitem[{G. Narayanan \& A. Banerjee(2021)Narayanan \&
  Banerjee}]{narayanan2021star}
Narayanan, G., \& Banerjee, A. 2021, \bibinfo{title}{Star Formation in
  Superthin Galaxies,} arXiv preprint arXiv:2104.04216

\bibitem[{G. Narayanan {et~al.}(2024)Narayanan, Banerjee,
  {et~al.}}]{narayanan2024does}
Narayanan, G., Banerjee, A., {et~al.} 2024, \bibinfo{title}{How does a low
  surface brightness galaxy form spiral arms?} arXiv preprint arXiv:2407.02916

\bibitem[{J.~F. Navarro {et~al.}(1997)Navarro, Frenk, \&
  White}]{navarro1997universal}
Navarro, J.~F., Frenk, C.~S., \& White, S.~D. 1997, \bibinfo{title}{A universal
  density profile from hierarchical clustering,} The Astrophysical Journal,
  490, 493

\bibitem[{L. Ossa-Fuentes {et~al.}(2023)Ossa-Fuentes, Borlaff, Beckman, Marcum,
  \& Fanelli}]{ossa2023flares}
Ossa-Fuentes, L., Borlaff, A.~S., Beckman, J.~E., Marcum, P.~M., \& Fanelli,
  M.~N. 2023, \bibinfo{title}{Flares, Warps, Truncations, and Satellite: The
  Ultra-thin Galaxy UGC 11859,} The Astrophysical Journal, 951, 149

\bibitem[{L.~E. P{\'e}rez-Monta{\~n}o {et~al.}(2022)P{\'e}rez-Monta{\~n}o,
  Rodriguez-Gomez, Cervantes~Sodi, Zhu, Pillepich, Vogelsberger, \&
  Hernquist}]{perez2022formation}
P{\'e}rez-Monta{\~n}o, L.~E., Rodriguez-Gomez, V., Cervantes~Sodi, B., {et~al.}
  2022, \bibinfo{title}{The formation of low surface brightness galaxies in the
  IllustrisTNG simulation,} Monthly Notices of the Royal Astronomical Society,
  514, 5840

\bibitem[{V. Perret(2016)Perret}]{perret2016dice}
Perret, V. 2016, \bibinfo{title}{DICE: Disk Initial Conditions Environment,}
  Astrophysics Source Code Library, ascl

\bibitem[{A. Pillepich {et~al.}(2018)Pillepich, Springel, Nelson, Genel,
  Naiman, Pakmor, Hernquist, Torrey, Vogelsberger, Weinberger,
  {et~al.}}]{pillepich2018simulating}
Pillepich, A., Springel, V., Nelson, D., {et~al.} 2018,
  \bibinfo{title}{Simulating galaxy formation with the IllustrisTNG model,}
  Monthly Notices of the Royal Astronomical Society, 473, 4077

\bibitem[{K.~C. Ramirez {et~al.}(2025)Ramirez, Sodi, Rosas-Guevara,
  P{\'e}rez-Monta{\~n}o, \& Bonoli}]{ramirez2025study}
Ramirez, K.~C., Sodi, B.~C., Rosas-Guevara, Y., P{\'e}rez-Monta{\~n}o, L.~E.,
  \& Bonoli, S. 2025, \bibinfo{title}{Study of barred galaxies in
  IllustrisTNG100: the case of low surface brightness galaxies,} arXiv preprint
  arXiv:2504.02145

\bibitem[{A.~B. Romeo \& N. Falstad(2013)Romeo \& Falstad}]{romeo2013simple}
Romeo, A.~B., \& Falstad, N. 2013, \bibinfo{title}{A simple and accurate
  approximation for the Q stability parameter in multicomponent and
  realistically thick discs,} Monthly Notices of the Royal Astronomical
  Society, 433, 1389

\bibitem[{K. Saha(2014)Saha}]{saha2014disc}
Saha, K. 2014, \bibinfo{title}{Disc heating: possible link between weak bars
  and superthin galaxies,} arXiv preprint arXiv:1403.1711

\bibitem[{K. Saha {et~al.}(2010)Saha, Tseng, \& Taam}]{saha2010effect}
Saha, K., Tseng, Y.-H., \& Taam, R.~E. 2010, \bibinfo{title}{The effect of bars
  and transient spirals on the vertical heating in disk galaxies,} The
  Astrophysical Journal, 721, 1878

\bibitem[{J. {Schaye} {et~al.}(2015){Schaye}, {Crain}, {Bower}, {Furlong},
  {Schaller}, {Theuns}, {Dalla Vecchia}, {Frenk}, {McCarthy}, {Helly},
  {Jenkins}, {Rosas-Guevara}, {White}, {Baes}, {Booth}, {Camps}, {Navarro},
  {Qu}, {Rahmati}, {Sawala}, {Thomas}, \& {Trayford}}]{2015MNRAS.446..521S}
{Schaye}, J., {Crain}, R.~A., {Bower}, R.~G., {et~al.} 2015,
  \bibinfo{title}{{The EAGLE project: simulating the evolution and assembly of
  galaxies and their environments},} Monthly Notices of the Royal Astronomical
  Society, 446, 521, \dodoi{10.1093/mnras/stu2058}

\bibitem[{J. Sellwood(2013)Sellwood}]{sellwood2013relaxation}
Sellwood, J. 2013, \bibinfo{title}{Relaxation in N-body simulations of disk
  galaxies,} The Astrophysical Journal Letters, 769, L24

\bibitem[{J. Sellwood(2016)Sellwood}]{sellwood2016bar}
Sellwood, J. 2016, \bibinfo{title}{Bar instability in disk--halo systems,} The
  Astrophysical Journal, 819, 92

\bibitem[{J. Sellwood \& O. Gerhard(2020)Sellwood \&
  Gerhard}]{sellwood2020three}
Sellwood, J., \& Gerhard, O. 2020, \bibinfo{title}{Three mechanisms for bar
  thickening,} Monthly Notices of the Royal Astronomical Society, 495, 3175

\bibitem[{J.~A. {Sellwood} \& N.~W. {Evans}(2001){Sellwood} \&
  {Evans}}]{2001ApJ...546..176S}
{Sellwood}, J.~A., \& {Evans}, N.~W. 2001, \bibinfo{title}{{The Stability of
  Disks in Cusped Potentials},} \apj, 546, 176, \dodoi{10.1086/318228}

\bibitem[{S. Sharma {et~al.}(2014)Sharma, Bland-Hawthorn, Binney, Freeman,
  Steinmetz, Boeche, Bienayme, Gibson, Gilmore, Grebel,
  {et~al.}}]{sharma2014kinematic}
Sharma, S., Bland-Hawthorn, J., Binney, J., {et~al.} 2014,
  \bibinfo{title}{Kinematic modeling of the Milky Way using the RAVE and GCS
  stellar surveys,} The Astrophysical Journal, 793, 51

\bibitem[{B.~C. Sodi \& O.~S. Garc{\'\i}a(2017)Sodi \&
  Garc{\'\i}a}]{sodi2017low}
Sodi, B.~C., \& Garc{\'\i}a, O.~S. 2017, \bibinfo{title}{Do Low Surface
  Brightness Galaxies Host Stellar Bars?} The Astrophysical Journal, 847, 37

\bibitem[{D. Sotillo-Ramos {et~al.}(2023)Sotillo-Ramos, Donnari, Pillepich,
  Frankel, Nelson, Springel, \& Hernquist}]{sotillo2023disc}
Sotillo-Ramos, D., Donnari, M., Pillepich, A., {et~al.} 2023,
  \bibinfo{title}{Disc flaring with TNG50: diversity across Milky Way and M31
  analogues,} Monthly Notices of the Royal Astronomical Society, 523, 3915

\bibitem[{A. Toomre(1964)Toomre}]{toomre1964gravitational}
Toomre, A. 1964, \bibinfo{title}{On the gravitational stability of a disk of
  stars,} The Astrophysical Journal, 139, 1217

\bibitem[{A. Toomre(1981)Toomre}]{toomre1981amplifies}
Toomre, A. 1981, in Structure and evolution of normal Galaxies, 111--136

\bibitem[{M. Vogelsberger {et~al.}(2014)Vogelsberger, Genel, Springel, Torrey,
  Sijacki, Xu, Snyder, Nelson, \& Hernquist}]{vogelsberger2014introducing}
Vogelsberger, M., Genel, S., Springel, V., {et~al.} 2014,
  \bibinfo{title}{Introducing the Illustris Project: simulating the coevolution
  of dark and visible matter in the Universe,} Monthly Notices of the Royal
  Astronomical Society, 444, 1518

\bibitem[{R. Weinberger {et~al.}(2020)Weinberger, Springel, \&
  Pakmor}]{weinberger2020arepo}
Weinberger, R., Springel, V., \& Pakmor, R. 2020, \bibinfo{title}{The Arepo
  public code release,} The Astrophysical Journal Supplement Series, 248, 32

\bibitem[{D. Xu {et~al.}(2024)Xu, Gao, Bottrell, Yesuf, \&
  Shi}]{xu2024illustristng}
Xu, D., Gao, H., Bottrell, C., Yesuf, H.~M., \& Shi, J. 2024,
  \bibinfo{title}{IllustrisTNG in the HSC-SSP: No Shortage of Thin Disk
  Galaxies in TNG50,} The Astrophysical Journal, 974, 88

\bibitem[{Z. Zhou {et~al.}(2021)Zhou, Ma, \& Wu}]{zhou2021correlation}
Zhou, Z., Ma, J., \& Wu, H. 2021, \bibinfo{title}{On the Correlation between
  Atomic Gas and Bars in Galaxies,} The Astronomical Journal, 161, 260

\end{thebibliography}

\end{document}